\documentclass[a4paper]{jpconf}
\usepackage[pdftex]{graphicx}

\begin{document}

\title{The Run Control system of the NA62 experiment}

\author{Nicolas Lurkin$^{1,}$\footnote[2]{On behalf of the NA62 collaboration: Birmingham, Bratislava, Bristol, Bucharest, CERN, Dubna, Fairfax, Ferrara, Florence, Frascati, Glasgow, Liverpool, Louvain, Mainz, Merced, Moscow, Naples, Perugia, Pisa, Prague, Protvino, Rome I, Rome II, San Luis Potosí, Sofia, TRIUMF, Turin, Vancouver.}{\rm , supported by ERC Starting Grant 336581}}
{\address{$^1$ School of Physics and Astronomy, University of Birmingham, Birmingham B15 2TT, UK}}
\ead{ncl@hep.ph.bham.ac.uk}

\begin{abstract}
The NA62 experiment at the CERN SPS aims at measuring the ultra-rare decay 
$K^+ \to \pi^+ \nu \bar{\nu}$ with 10 \% accuracy.  This can be achieved by detecting about 100 Standard Model events with 10\% background in 2--3 years of data taking.
The experiment consists of a large number of subsystems dedicated to the detection of the incoming kaon and  outgoing pion and also  focusing on particle identification and vetoing capabilities. 
Run Control has been designed to link their trigger and data acquisition systems in a single central application easily controllable by educated but non-expert operators. 
The application has been continuously evolving over time, integrating new equipments and taking into account  requirements and feedback from operation. 
Future development includes  a more automatized system  integrating the knowledge acquired during the operation of the experiment.
\end{abstract}

\section{The NA62 experiment}
The NA62 experiment at the CERN SPS aims to collect about 100 standard model (SM) events  of the ultra-rare kaon decay $K^+ \to \pi^+ \nu \bar{\nu}$ with less than 10\% background in 2--3 years of data taking. 
This process is forbidden at tree level and can occur only through a Flavour Changing Neutral Current loop. 
It is theoretically very clean 
and the prediction of the branching ratio  has been computed within the SM  \cite{Buras_2015} to an exceptionally high degree of precision: ${\mathcal{B}(K^+\to\pi^+\nu\bar{\nu})_{th} = (8.4 \pm 1.0) 10^{-11}}$. 
This channel is therefore an excellent probe of the Standard Model  and any deviation from this value would indicate a new physics process.
A total of 7 events were observed by  the  BNL E787/E949 experiments \cite{E787/E949} including an estimated background of ($2.6  \pm 0.4$) events.
The measured branching ratio ${\mathcal{B}(K^+\to\pi^+\nu\bar{\nu})_{exp}=(1.73^{+1.15}_{-1.05})\times 10^{-10}}$ has an uncertainty by far too large to conclude on a possible deviation from the SM predictions known at $\sim$ 10\% precision.

To achieve its ambitious goal, NA62 needs to accumulate $\cal{O}$$(10^{13})~K^+$ decays using a secondary 75 GeV/c momentum unseparated hadron beam containing only  $\sim$6\% $K^+$.  
Many different sub-detectors have been developed to  identify the signal process and fight the 
background from other kaon decays and  beam accidental coincidences.  The current status and achieved sensitivity of the experiment is reported in \cite{GRkaon16}.

\section{The NA62 Run Control}
Run Control is an application that centrally controls and monitors all equipment involved in the trigger and data acquisition (TDAQ) system.
Its purpose is to allow a non-expert operator to supervise the data taking easily while still giving specialists the possibility to achieve a high level of control of their own equipment.

\paragraph{\bf Technologies and architecture:}
Operation of the experiment involves other  control systems dedicated to hardware equipment (Dectector Control System DCS, Gas System, Cryogenics). A coherent and natural choice was 
 the industrial software ``WinCC Open Architecture".
This software is the central part of two frameworks - JCOP \cite{JCOP} and UNICOS \cite{UNICOS}- already widely used at CERN in the LHC experiments. 
Some of the components provided by these developer toolkits are of special relevance for NA62:
\begin{itemize}
  \item DIM (Distributed Interface Management) is the communication layer between Run Control and the numerous equipment distributed across the experiment \cite{DIM};
  \item The FSM toolkit manages the  finite state machines with SMI++ processes \cite{SMI++} according to their definition (states, transition rules, actions);
  \item The configuration database tool defines and/or applies sets of parameters from a database;
  \item The Farm Monitoring and Control  manages the PC farm which receives information from individual subdetectors and builds the complete events.
\end{itemize}

The NA62 TDAQ system is composed of many devices, most of them operating 
in a specific way.
This complication is overcome by internally using a Finite State Machine (FSM) model of each device, therefore presenting a simple and common control to the operator.
Little equipment-specific knowledge is integrated in Run Control, while a common generic-device interface is created to transmit commands and information.
Simple basic instructions are sent through this channel, and are received by the device control software that is responsible for executing the proper sequence of actions on the hardware.
More complex configurations rely on a scheme of flexible XML files where the details of the configuration are again handled by the receiver.
Multiple sets of configurations are stored in an Oracle database and can be quickly loaded and distributed to the relevant equipment when needed. 
This architecture allows Run Control and the controlled systems to evolve independently from each other while keeping a permanent compatibility. 

The software is implemented as a hierarchical tree of FSM where the state of each node is either defined by a set of rules summarizing the state of the child-nodes or by the evaluation of logical expressions involving parameters transmitted by the devices.

The external nodes of the tree implement the FSM models representing the specific hardware or software devices (boards, computer, crates and control softwares). 
Their states are evaluated according to the value of parameters received from the corresponding piece of equipment. 
The internal nodes represent logical subsystems (subdetectors, PC farm), summarizing the states of their own child-nodes following a set of rules. 
Finally the top node further aggregates the state of the subsystems to represent the global state of the experiment.

\begin{figure}
\centering
\includegraphics[width=0.55\linewidth]{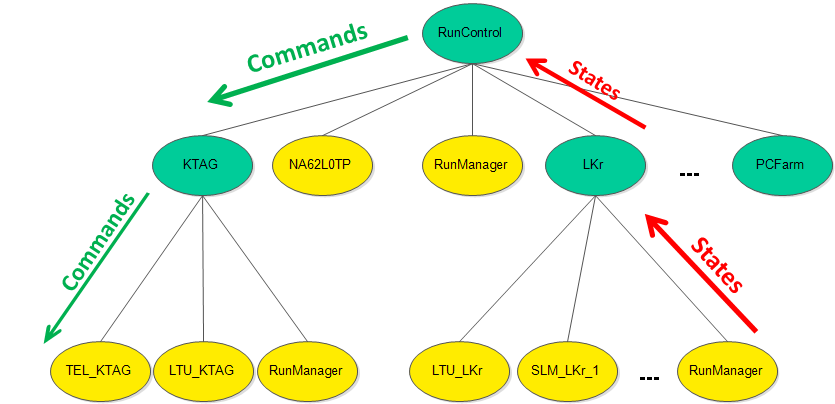}
\caption{Tree-like hierarchy of units. Green units are logical units and yellow are device units. Commands are
propagated downwards and states are propagated upwards and summarized in the logical units.}
\label{FSMTree}
\end{figure}

A change of state is always propagated upwards from the device where this change originates to the top node. 
Conversely, a command can be issued at any node level and is always propagated downwards to all the child-nodes until it reaches a device node (Figure \ref{FSMTree}).
At this point, the command is generated and transmitted to the standardized interface through the network using the DIM protocol. 

\paragraph{\bf Infrastructure}
Run Control is a distributed system spread across different machines in the experiment. 
The core of the system is located on a dedicated WinCC OA data server on the technical network. 
A machine bridging the networks hosts the DIM managers which are responsible for loosely binding all devices in the experiment and Run Control. 
The experiment DAQ continues to run correctly standalone in case of a connectivity problem or if this node is cut from one of the  networks. 
As soon as communication is re-established, all clients reconnect to the DIM servers and Run Control resumes control over the experiment. 
The bridge also receives external information 
from the beam line complex.
The user interface itself runs on a different computer in the control room and is remotely connected to the main system.

\paragraph{\bf Future developments}
The first version of Run Control was successfully deployed for a dry run in 2012 and has continuously evolved since then.
New equipment and subdetectors have been delivered and subsequently integrated.
The experience acquired during the following years and the feedback from the operators brought Run Control to a good level of reliability and usability.

The next steps in the development are the automatization of known procedures and automatic error detection and recovery.
The ELectronic Eye of NA62 (ELENA62) is the core system being developed and already partially deployed. 
It provides a framework for monitoring functions, interactions with the operator through visual and audio notifications, and confirmation windows.
It is highly customizable and configurable as the monitoring and control of specific components of the experiment are provided through plug-in modules.

Among those already  in use to increase the efficiency and quality of the data taking are:
the PC farm module that handles the PC nodes by restarting the acquisition software after a crash is detected, or that reboots the nodes when a hardware error occurs;
two monitoring modules of the beam-line magnets and vacuum in the decay region that provide a fast feedback to the operator;
another module  that automatizes the start and end of run procedure, additionally guiding the operator when a manual action is necessary.  

\paragraph{ }
In conclusion, Run Control has already proved its usability and good reliability, confirming the technological choices made for its design.
It has now reached a matured level of development with minimal  daily maintenance, allowing the integration of the experience acquired during its operation into a more autonomous system.

\end{document}